\documentclass{iopart}

\usepackage{graphicx}
\usepackage{dcolumn}
\usepackage{bm}
\usepackage{color}
\providecommand{\eqref}[1]{(\ref{#1})}

\begin{document}

\topical[Optical spin pumping of QWs]{Time-resolved and
continuous-wave optical spin pumping of semiconductor quantum
wells }

\author{G.~V.~Astakhov$^{1,2}$, M.~M.~Glazov$^{2}$, D.~R.~Yakovlev$^{2,3}$, E.~A.~Zhukov$^{3,4}$,
W.~Ossau$^{1}$, L.~W.~Molenkamp$^{1}$, and M.~Bayer$^3$}
\address{$^1$Physikalisches Institut (EP3),
Universit\"{a}t W\"{u}rzburg, D-97074 W\"{u}rzburg, Germany\\
$^2$A.F. Ioffe Physico-Technical Institute, Russian Academy of
Sciences, 194021 St. Petersburg, Russia\\
 $^3$Experimentelle Physik 2, Technische Universit\"at Dortmund,
D-44221 Dortmund, Germany\\
$^4$Faculty of Physics, M.V. Lomonosov Moscow State University,
119992 Moscow, Russia}

\begin{abstract}

Experimental and theoretical studies of all-optical spin pump and
probe of resident electrons in CdTe/(Cd,Mg)Te semiconductor
quantum wells are reported. A two-color Hanle-MOKE technique
(based on continuous-wave excitation) and time-resolved Kerr
rotation in the regime of resonant spin amplification (based on
pulsed excitation) provide a complementary measure of electron
spin relaxation time. Influence of electron localization on
long-lived spin coherence is examined by means of spectral and
temperature dependencies. Various scenarios of spin polarization
generation (via the trion and exciton states) are analyzed and
difference between continuous-wave and pulsed excitations is
considered. Effects related to inhomogeneous distribution of
$g$-factor and anisotropic spin relaxation time on measured
quantities are discussed.

\end{abstract}



\section{Introduction}

\begin{figure}[btp]
\includegraphics[width=0.7\textwidth]{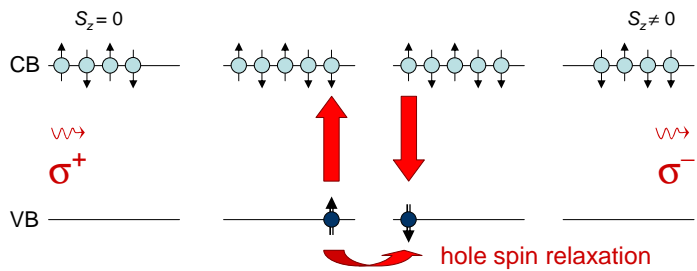}
\caption{Schematic illustration of the optical spin pumping of
resident electrons in bulk semiconductors. Wavy lines show
incident $\sigma^+$ and emitted $\sigma^-$ photons. Arrows
$\uparrow$ ($\downarrow$) correspond to the electrons with $s_z =
+ 1/2$ ($s_z = - 1/2$). Arrows $\Uparrow$ ($\Downarrow$)
correspond to the heavy-holes with $j_z = + 3/2$ ($j_z = - 3/2$).
$S_z$ denotes the total electron spin of electrons.}
\label{fig1_scheme}
\end{figure}

Optical spin pumping of resident electrons in bulk semiconductors
has been established since the seventies of the past
century~\cite{Book_OO}. The circularly polarized light induces
interband transitions between the hole states in the valence band
(VB) and the electron states ($e$) in the conduction band (CB).
Owing to the optical selection rules the allowed photo-generated
electron-hole pairs in the case of $\sigma^{+}$ excitation are
$(hh,+3/2;e,-1/2)$ and $(lh,+1/2;e,+1/2)$. Here, $(hh,\pm 3/2)$
denotes the heavy-hole state with spin projection $j_z = \pm 3/2$
on the incident direction of the light, $(lh,\pm 1/2)$ denotes the
light-hole state with $j_z = \pm 1/2$ and $(e,\pm 1/2)$ the
electron state with $s_z = \pm 1/2$. Since the matrix elements of
the heavy-hole and light-hole transitions relate as $\sqrt{3}$ to
1, $\sigma^{+}$ excitation creates predominantly spin-down
electrons, i.e., with $s_z = -1/2$ (Fig.~\ref{fig1_scheme}). If
hole spin flip occurs, the hole can recombine with a spin-up
electron emitting $\sigma^{-}$-polarized photon. As a result, the
total electron spin $S_z = (n_{+}-n_{-}) / 2$ after recombination
is not zero any more ($n_{\pm}$ is concentration of $s_z = \pm
1/2$ electrons). The net spin polarization is equal
\begin{equation}\label{Sz-rho}
\rho_e = 2 S_z/n_e \,,
\end{equation}
where $n_e$ is the total electron concentration. Typically, a hole
looses its spin quickly and the electron spin relaxation time
$\tau_s$ is rather long. As a consequence, the optically created
electron spins are accumulated in $s_z = -1/2$ state under
continuous wave (cw) excitation with $\sigma^{+}$ polarization.
This implies optical spin pumping. The theoretical limit of the
steady-state net spin polarization is $\rho_e = 0.5$
\cite{Book_OO}. This is valid for bulk semiconductors with
degenerated light-hole and heavy-hole states, like GaAs and CdTe.

In bulk $n$-type semiconductors the circular polarization degree
of the edge photoluminescence (PL)
${P=(I_{{+{}}}^{{+{}}}-I_{{+{}}}^{{-{}}})/(I_{{+{}}}^{{+{}}}+I_{{+{}}}^{{-{}}})}$
is proportional to the net spin polarization $P = \rho_e / 2$.
Here, ${I_{{+{}}}^{{\pm {}}}}$  is the PL intensity of the
corresponding optical transition under  ${\mathit{{\sigma}}^{{\pm
{}}}}$ excitation detected in
${\mathit{{\sigma}}^{{+{}}}}$-polarization component. When an
in-plane magnetic field $B$ is applied electron spins start to
precess around magnetic field direction with the Larmor frequency
$\omega_L=g_{{e}}\mu_{{B}}B/\hbar$, where $g_e$ is electron
$g$-factor and $\mu_B$ is the Bohr magneton. As a result, $S_z$
decreases following
\begin{equation}\label{eq1}
{S_{{z}}\left(B\right)=\frac{S_{{z}}\left(0\right)}{1+(B/{B_{{1/2}}})^{{2}}}}.
\end{equation}
Such a behavior detected by means of the PL circular polarization
$P \propto S_z$ is known as the Hanle effect \cite{Book_OO}. The
average electron spin drops by a factor of $2$ when the electron
spin lifetime $T_s$ and the Larmor frequency $\omega_L$ satisfy
the condition ${\mathit{{\omega}}_{{L}}T_{{s}}=1}$. This condition
defines characteristic magnetic field ${B_{{1/2}}=\hbar
/(\mathit{{\mu}}_{{B}}g_{{e}}T_{{s}})}$  allowing to obtain
${T_{{s}}}$ if $g$-factor is known. These experiments can be
performed under cw optical excitation.

In $n$-type semiconductors an additional spin decay channel is the
recombination of resident electrons with photogenerated holes. As
a consequence the spin lifetime depends on pump density $W$ as
\begin{equation}\label{eq1:W}
T_s^{-1} = \tau_s^{-1} (1 +W/W_0) \,.
\end{equation}
Using Eq.~(\ref{eq1}) in the limit of low excitation density
(i.e., $W$ much below the characteristic pump density $W_0$) one
can find the spin relaxation time of `unperturbed' electron system
$\tau_s$, as $T_s \rightarrow \tau_s$.

Time resolved spectroscopy based on pulsed photoexcitation offers
straightforward access to the carrier spin dynamics (for overview
see Ref.~\cite{Book_Spin}). For example, polarized
photoluminescence measured by streak-camera allows to follow spin
relaxation during the exciton lifetime. Faraday and Kerr rotation
techniques can monitor magnetization induced by spin polarized
carriers at time delays not limited by exciton recombination. In
external magnetic fields not only spin dynamics but also Zeeman
splitting for carriers and excitons can be measured from the
period of signal oscillations. Respectively, the direct evaluation
of carrier and exciton $g$-factors became possible. Partial
limitations of time-resolved techniques are related to relatively
high peak density of photoexcitation as compared to cw excitation.
This may cause perturbation of electronic system (carrier heating,
nonequilibrium distribution) which can affect the spin dynamics.

In low-dimensional quantum well (QW) heterostructures the optical
spin pumping can be modified due to following reasons. (i) Quantum
confinement and strain lift up the degeneracy of light-hole and
heavy-hole states. Hence, their selective excitation becomes
possible and theoretical limit for the net spin polarization in
this case is increased to 100\% ($| \rho_e | = 1$). (ii) In
$n$-type modulation doped QWs a negatively charged trion (T)
becomes stable. It is composed of a hole and two electrons, the latter
form the spin singlet~\cite{T_rev}. The singlet state itself does
not contribute to the spin polarization, but resonant excitation
of trions and/or trion formation from excitons provide an
efficient mechanism for polarization of resident electrons in QWs
\cite{zhukov07}.

The goal of this paper is to perform a comparative study of
electron spin dynamics under cw and pulsed photoexcitation and to
clarify whether the same values of spin relaxation times can be
achieved by both techniques. Experiments are done for the very
same samples, which are CdTe/(Cd,Mg)Te quantum well structures
with diluted concentration of resident electrons.

The paper is organized as follows. We start with model
considerations of the generation of spin polarization and spin
coherence in quantum wells with diluted two-dimensional electron
gas. Then we present experimental results for cw and pulsed
excitation, analyze them and conclude.


\section{Theory}

In the present section different scenarios of spin polarization
generation and spin dynamics in transverse magnetic field are
discussed. We consider both continuous wave (cw) and pulsed
excitation regimes and we discuss one after another the mechanisms
of spin pumping under excitation at the trion and exciton
resonances.

\subsection{Spin coherence excitation}\label{exc}

Here we analyze the processes responsible for the generation of
the electron spin coherence in QWs with low density
two-dimensional electron gas (2DEG). It is assumed that the
temperature measured in energy units $k_B T$ is smaller than both
exciton and trion binding energies and that the resident electron
concentration is so small that trion and exciton states are stable
(i.e. these states are not screened by electron-electron
interactions). More detailed analysis is published in
\cite{zhukov07}.

\subsubsection{Trion resonant pumping}\label{tr}

\paragraph{Continuous wave excitation.}

\begin{figure}[tbp]
\includegraphics[width=0.7\textwidth]{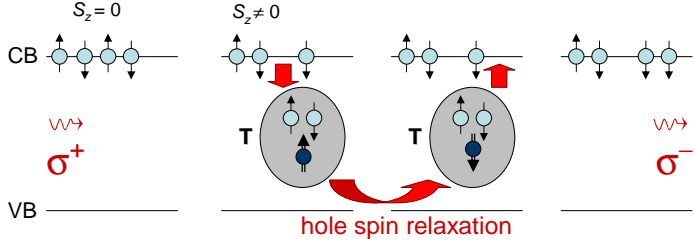}
\caption{Schematic illustration of the optical spin pumping of
resident electrons in QWs, resonant excitation to the trion
(T).}\label{fig:illustr}
\end{figure}

First, we consider the case where the trion is resonantly excited.
The trion ground state forms the singlet spin structure, i.e., the
spins of the two electrons involved in the trion are oriented
antiparallel \cite{4,4_ast}. A $\sigma^+$ polarized photon creates
an electron-hole pair $(hh,+3/2;e,-1/2)$ and simultaneously an
electron with the spin projection $s_z = +1/2$ is picked out from
the resident ensemble in order to form a trion
(Fig.~\ref{fig:illustr}). Thus, the resident ensemble becomes
depleted of electrons with the spin projection opposite to that of
photocreated ones, i.e. in our case the number of spin $s_z =
+1/2$ electrons decreases. This means that a spin polarization of
resident carriers $S_z \neq 0$ appears. The electrons returning to
the ensemble after trion recombination can compensate this
polarization. However, if the hole spin relaxation time in the
trion is much shorter than the trion radiative lifetime the
returned electrons are spin unpolarized and the spin polarization
generated at the moment of absorption is conserved.

From the considerations above it is clear that at the initial
moment each absorbed $\sigma^{\pm}$ photon increases the spin of
resident electrons by $\mp 1/2$. In order to study the pump power
dependence of the spin polarization generation we use the system
of coupled kinetic equations describing the populations of
electrons with different spins $n_{\pm}$, and populations of
singlet trions $T_\pm$ (with the heavy-hole spin $\pm 3/2$)
\begin{eqnarray}
&& -\frac{T_+}{\tau_0^T} - \frac{T_+}{2\tau_s^T} + \frac{T_-}{2\tau_s^T} + n_+ G = 0 ,\label{steady}\\
&& -\frac{T_-}{\tau_0^T} - \frac{T_-}{2\tau_s^T} + \frac{T_+}{2\tau_s^T} = 0 ,\nonumber\\
&& -\frac{n_+}{2\tau_s} + \frac{n_-}{2\tau_s} + \frac{T_+}{\tau_0^T} - n_+ G = 0 ,\nonumber\\
&& -\frac{n_-}{2\tau_s}  +\frac{n_-}{2\tau_s} +
\frac{T_-}{\tau_0^T} = 0 \nonumber.
\end{eqnarray}
Here it is assumed that the pumping is $\sigma^+$ polarized, $G$
is the generation rate (being proportional to the pump density
$W$), $\tau^T_0$ is the trion radiative lifetime, $\tau^T_s$ is
the spin relaxation time of a hole in a trion (trion spin
relaxation time), and $\tau_s$ is the spin relaxation time of
resident electrons. System \eqref{steady} should be complemented
with the condition
\[
 T_+ + T_- + n_+ + n_-=n_e,
\]
which describes conservation of resident electrons $n_e$.

\begin{figure}[tbp]
\includegraphics[width=0.6\textwidth]{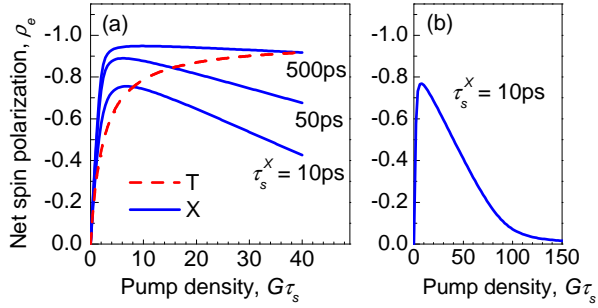}
\caption{(a) Spin polarization of resident electrons $\rho_e$ as a
function of the generation rate in units of $G \tau_s$ under
$\sigma^+$ excitation. Dashed curve corresponds to the trion
resonant pumping. Solid curves correspond to the exciton resonant
pumping. They are labeled by the respective electron-in-exciton
spin relaxation time, $\tau_s^X$. All curves are obtained for the
electron spin relaxation time $\tau_s = 2$~ns, the exciton and
trion radiative lifetimes $\tau_0^X = \tau_0^T = 50$~ps, the spin
relaxation time of a hole in a trion $\tau_s^T = 10$~ps. The
exciton to trion conversion coefficient is $\gamma n_e = 1$~ps.
(b) The spin polarization of resident electrons (under the exciton
resonant pumping) shown for a wide range of $G\tau_s$ ($\tau_s^X =
10$~ps).}\label{fig6}
\end{figure}

The total steady state spin of resident electrons can be written
as
\begin{equation}\label{gen_spin}
S_z = \frac{n_+ - n_-}{2} = -\frac{n_e}{2} \frac{G \tau_s }{2
\tau_s^T / \tau^T + G \tau_0^T\tau_s^T/\tilde \tau},
\end{equation}
where $\tau^T = \tau_s^T\tau_0^T/(\tau_s^T+\tau_0^T)$ is the trion
spin lifetime including both the trion lifetime $\tau_0^T$ and the
trion spin relaxation time $\tau_s^T$, and
\[
\frac{1}{\tilde \tau}= \frac{1}{\tau^T} + \frac{\tau_s}{\tau_s^T
\tau_0^T}.
\]
In experimental conditions \cite{zhukov07} the electron spin
relaxation time $\tau_s\gg \tau_0^T,\tau_s^T$. Therefore
Eq.~\eqref{gen_spin} can be recast as
\begin{equation}\label{gen_spin_1}
S_z = \frac{n_+ - n_-}{2} = -\frac{n_e}{2} \frac{G \tau_s }{2
\tau_s^T / \tau^T + G \tau_s}.
\end{equation}
The net spin polarization $\rho_e = 2 S_z/n_e$ is plotted as a
function of $G\tau_s$ in Fig.~\ref{fig6}(a) by the dashed line.
The parameters of calculation are given in the figure caption.

At low pumping densities $G\tau_s \ll \tau_s^T/\tau^T$, i.e.,
where the spin polarization is small and no saturation effects are
observed the electron spin grows linearly with the pumping density
as
\begin{equation}\label{low}
 S_z = -\frac{n_e}{4} G \frac{\tau_s\tau^T}{\tau_s^T}.
\end{equation}
It is worth noting that the absolute value of the electron spin
$|S_z|$ decreases with an increase of trion spin relaxation time
$\tau_s^T$. It is in agreement with the qualitative picture
outlined above: in order to induce the spin coherence of resident
electrons the spin-flip of a hole in a trion is required. That
means that $\tau_s^T$ should be comparable or shorter than the
trion lifetime $\tau_0^T = 50$~ps. With further increase of the
pump density the total spin of resident electrons grows
sublinearly (Fig.~\ref{fig6}). This is because the larger $|S_z|$
the less resident electrons with proper spin required to form the
trion (i.e., $s_z = +1/2$) are left.

At high pump densities $G\tau_s \gg \tau_s^T/\tau^T$ the electron
spin saturates at
\begin{equation}\label{satur:spin:tr}
S_z = -\frac{n_e}{2} \frac{\tau_s \tilde \tau}{\tau_0^T \tau_s^T}
\approx -\frac{n_e}{2}.
\end{equation}
This means that almost all resident electrons become spin
polarized: $\rho_e \approx -1$. The negative sign results from the
$\sigma^+$ pumping when electrons with spin $s_z = -1/2$ are
created. The minority of electrons $s_z = +1/2$ are bound in
trions whose density is $n_e(\tau_0^T+\tau_s^T)/\tau_s \ll n_e$.
This implies that efficient spin pumping of resident electrons may
be achieved at relatively low excitation densities. Such an
unexpected behaviour is a direct consequence of the long spin
relaxation of resident carriers. With increasing pump density the
spin-up electrons ($s_z = +1/2$) are captured to the trions and
return back unpolarized. The stronger pumping the less electrons
have appropriate spin orientation and the number of trions
decreases.

\paragraph{Pulsed excitation.}

Under pulsed excitation the physical processes governing electron
polarization generation are essentially the same as for cw
excitation \cite{zhukov07}. After the trion formation the electron
gas becomes depleted of electrons with the spin projection
opposite to that of photocreated ones. The spin flip of a hole in
a trion results in the imbalance of the spins of resident
electrons and those returning after trion radiative recombination.

It is worth noting that the initial number of photogenerated
trions under pulsed resonant excitation ($n_0^T$) cannot exceed
$n_e/2$, where $n_e$ is the density of resident electrons. Thus,
the $n_0^T$ increases linearly with the pump intensity for small
excitation density and then saturates at the value $n_e/2$. The
total generated electron spin  at $\tau_0^T\gg \tau_s^T$ can be
evaluated as
\begin{equation}\label{trion_res_sat}
S_z = -\frac{n_e}{4}
G\tau_0^T/(1+G\tau_0^T),
\end{equation}
where $G$ is the generation rate being proportional to the pump
power.

It is instructive to compare this result with the result of
Eq.~\eqref{satur:spin:tr} obtained for the steady-state pumping.
In the case of pulsed excitation the maximum total spin of
resident electrons is $-n_e/4$, i.e., twice smaller than the
steady-state value.

\subsubsection{Exciton resonant pumping}\label{xx}

\paragraph{Continuous wave excitation.}

\begin{figure}[tbp]
\includegraphics[width=0.7\textwidth]{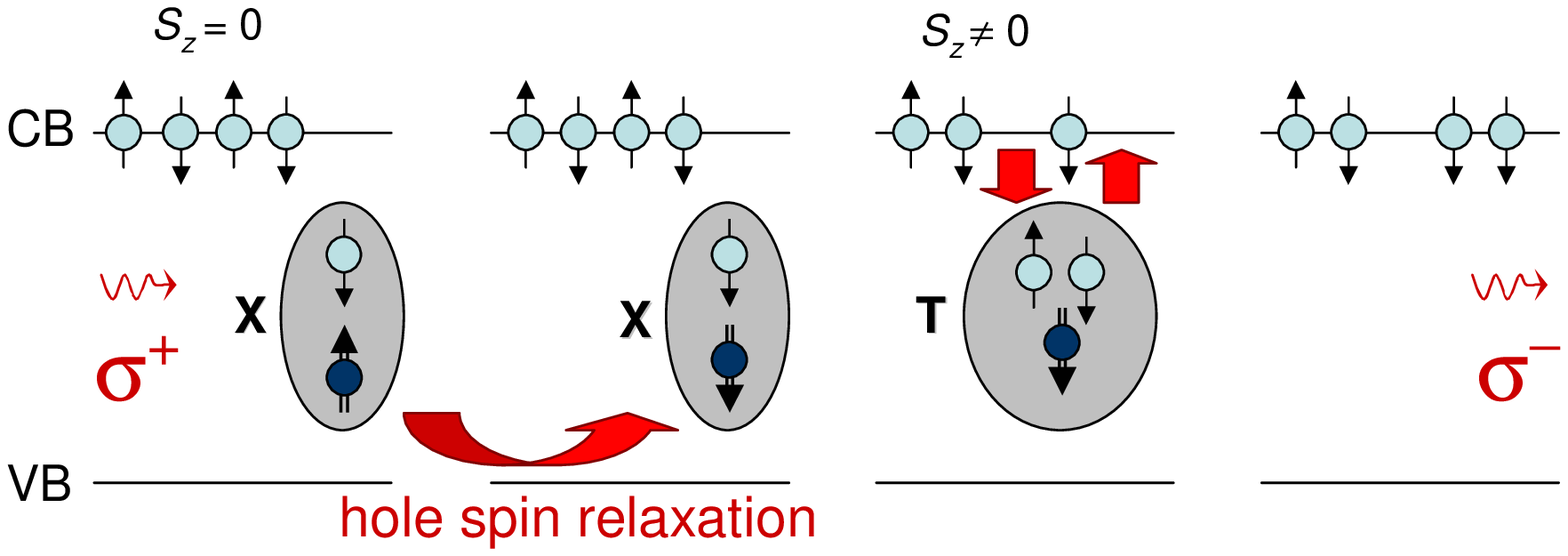}
\caption{Schematic illustration of the optical spin pumping of
resident electrons in QWs, resonant excitation to the exciton (X).
Limit of low pump density. }\label{fig:illusx}
\end{figure}

Next, we discuss the case of exciton resonant pumping.
Photocreated excitons can capture resident electrons and form
trions. This process is schematically shown in
Fig.~\ref{fig:illusx}. Under the assumption that the hole in the
exciton looses its spin rapidly one may label concentrations of
$X_+$ and $X_-$ excitons in accordance with the electron spin
projection in the exciton. The trions formed from these excitons
are unpolarized because the spin of the paired electrons in the
trion is zero and the hole spin is lost prior to trion formation.
The kinetic equations describing such a situation write
\begin{eqnarray}
&& -\frac{X_+}{\tau_0^X} - \frac{X_+}{2\tau_s^X} + \frac{X_-}{2\tau_s^X} - \gamma n_- X_+ = 0 \label{steady:exc}\\
&& -\frac{X_-}{\tau_0^X} - \frac{X_-}{2\tau_s^X} + \frac{X_+}{2\tau_s^X} - \gamma n_+ X_- + G_X = 0 \nonumber\\
&& -\frac{T}{\tau_0^T} + \gamma n_- X_+ + \gamma n_+ X_-= 0 ,\nonumber\\
&& -\frac{n_+}{2\tau_s} + \frac{n_-}{2\tau_s} + \frac{T}{2\tau_0^T} - \gamma n_+ X_- = 0 ,\nonumber\\
&& -\frac{n_-}{2\tau_s}  +\frac{n_-}{2\tau_s} +
\frac{T}{2\tau_0^T} - \gamma  n_- X_+ = 0 \nonumber.
\end{eqnarray}
Here $\tau_s^X$ is the electron-in-exciton spin relaxation time,
$\tau_0^X$ is the exciton radiative lifetime, $\gamma$ is a
constant describing the efficiency of trion formation from an
exciton, $T = T_+ + T_-$ is the total concentration of trions, and
$G_X$ is the exciton generation rate ($\sigma^+$ polarized pumping
is assumed).

\begin{figure}[tbp]
\includegraphics[width=0.7\textwidth]{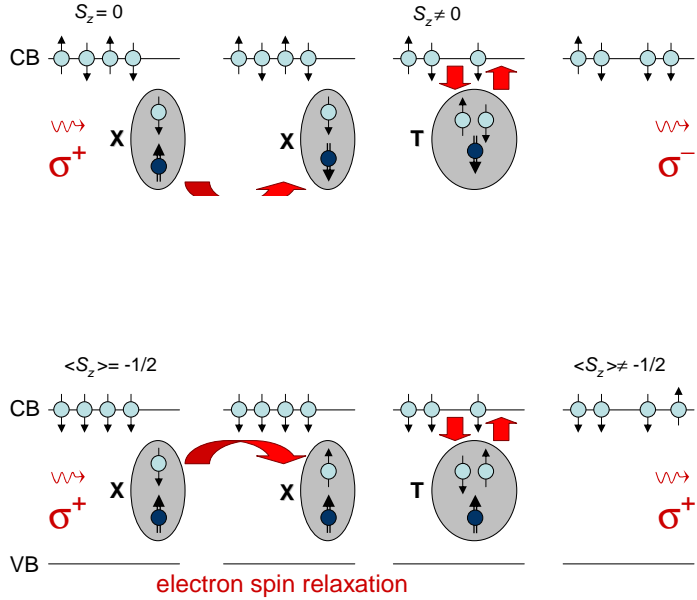}
\caption{Schematic illustration of the optical spin pumping of
resident electrons in QWs, resonant excitation to the exciton (X).
Limit of high pump density. }\label{fig:illusx_high}
\end{figure}

Solid lines in Fig.~\ref{fig6} show the spin of resident electrons
as a function of the pump density calculated for various values of
$\tau_s^X$. Contrary to the case of trion resonant excitation, an
initial increase of the net spin polarization is followed by its
decrease to zero for very high pump densities [shown in panel
(b)]. Qualitatively, such a behavior can be explained as follows.
In the absence of electron spin relaxation, the formation of
trions is prevented if the exciton concentration exceeds half of
the electron concentration. However, the spin flip on an
electron-in-exciton can provide the formation of trion even for
higher exciton concentrations (schematically shown in
Fig.~\ref{fig:illusx_high}). This is accompanied by a decrease of
the net spin polarization of electrons \cite{zhukov07}. At very
high pumping densities, $G_X\tau_s/n_e \gg 1$, all resident
electrons become bound to unpolarized trions and the system will
consist of $n_e$ unpolarized trions (no unbound electrons) and
$X=G_X\tau_0^X$ excitons with the total electron-in-exciton spin
$(X_+ - X_-)/2 = G_X \tau_0^X\tau_s^X/(\tau_0^X+\tau_s^X)$.

At low pump densities most of excitons form trions (the trion
formation process can be considered as the fastest one when
$\gamma n_e/2 \gg \tau_0^X,\tau_s^X$), which results in the spin
polarization of resident electrons. At $G_X\tau_s/n_e \ll 1$,
\begin{equation}\label{low:exc}
 S_z = -\frac{1}{2} G_X\tau_s. 
\end{equation}
and each absorbed photon participates in the trion formation
either directly or via the exciton state. Therefore, the
efficiency of the electron spin polarization is expected to be
nearly independent of the pump energy. Comparing
Eq.~\eqref{low:exc} and Eq.~\eqref{low} one can build a relation
between the trion generation rate $G$ entering Eq. \eqref{steady}
and exciton generation rate in Eq.~\eqref{steady:exc} as $G_X =
n_e G/2$. The spin polarization degree at high pump powers is
predicted to be strongly dependent on the excitation energy and
demonstrate qualitatively different behavior.

\paragraph{Pulsed excitation.}

In time-resolved experiments at low pump densities shortly after
the pulsed optical excitation all excitons are bound to trions and
the QW contains $n^X_0$ trions (where $n_0^X = G_X\tau_{pulse}$
with $\tau_{pulse}$ being the pulse duration, $\tau_{pulse} \ll
\tau_0^X$, is the number of photocreated excitons) and $n_e -
n^X_0$ resident electrons with a total spin
\begin{equation} \label{se_low}
|S_z| = n^X_0/2,    \quad n_0^X\leq n_e/2.
\end{equation}
As a result, $n^X_0$ spins of resident electrons contribute to the
Kerr rotation.

At higher excitation intensity ($n^X_0 \geq n_e/2$) all $n_e/2$
resident electrons with $s_z = + 1/2$ are bound to trions.
Therefore, in absence of electron-in-exciton spin relaxation
processes the trion density cannot exceed $n_e/2$, thus the total
spin density of the electron gas is limited by $|S_z| < n_e/4$.
The electron-in-exciton spin relaxation allows to convert the
remaining $n^X_0 - (n_e/2)$ excitons in trions. Obviously, the
maximum number of formed trions cannot exceed the concentration of
background electrons, $n_e$. The total spin of resident electrons
after the excitons and trions have recombined can be estimated as
(provided that the holes are unpolarized)
\begin{equation}\label{se_exc}
|{ S}_z|  \approx \frac{1}{4} \left\{
\begin{array}{cc}
n_e - \frac{2 n^X_0 - n_e}{1 + (2 \tau^X_s/\tau_0^X)} , & \mbox{if  }n_e
> \frac{2n^X_0 - n_e}{1 + (2 \tau^X_s/\tau_0^X)} \\
0, & \mbox{otherwise}.
\end{array}
\right.
\end{equation}
This equation is valid both for $B=0$ and $B \neq 0$ when $n^X_0
\geq n_e/2$, otherwise Eq. \eqref{se_low} holds. At $n^X_0 =
n_e/2$, the values of $|{ S}_z|$ given by Eqs.~(\ref{se_low}) and~(\ref{se_exc}) coincide and are equal to $n_e/4$. An initial
linear increase of $|S_z|$ followed by a linear decrease is seen
from Eq. \eqref{se_exc}.  The decrease of initial electron spin as
a function of pump intensity is steeper for smaller values of
$\tau_s^X/\tau_0^X$, i.e. for shorter hole spin relaxation times.
It is worth to stress that in this regime the electron spin
polarization vanishes at very high pumping whereas, under resonant
trion excitation, $|S_z|$ monotonously increases with increasing
pump power and saturates at $n_e/4$.

\subsection{Spin dynamics in transverse magnetic fields}\label{det}

Application of a transverse magnetic field (i.e., perpendicular to
the initial spin direction) induces electron spin precession about
the magnetic field, which is often called as spin beats.  In
general case time evolution of $S_z$ is described by
\begin{equation} \label{read2}
S_z(t) \sim {\rm e}^{- t/T_2^* } \cos{(\tilde{\Omega} t+\varphi)}
\end{equation}
where $\varphi$ is an initial phase and $T_2^*$ is the dephasing time
of electron spin ensemble. The dephasing time is contributed by
the coherence time of individual spins $T_2$ and by inhomogeneous
spin relaxation time $T_2^{inh}$ caused by, e.g., variation in
electron g-factors:  $1/T_2^*=1/T_2+1/T_2^{inh}$ (see Chapter 6 in
\cite{Book_Spin}). In the simplest case of isotropic and
homogeneous system the spin relaxation of resident electrons is
characterized by a single time constant $\tau_s=T_2$. In this case
$T_2^* = \tau_s$, and the frequency of spin beats is equal to the
Larmor frequency $\tilde{\Omega} = \omega_L$. In case of QWs
experimental behavior appears to be somewhat more complicated for
few reasons: (i) spin relaxation can be anisotropic
\cite{averkiev06}, (ii) the main mechanism of spin dephasing can
be caused by $g$-factor inhomogeneous distribution, (iii) the
initial phase can depend on magnetic field.

\subsubsection{Continuous wave excitation}\label{cwExcitation}

In many cases the regime of cw spin pumping in transverse magnetic
fields (and the resultant Hanle effect) may be described in terms
of the spin relaxation time $\tau_s$. The corrections due to
anisotropy and inhomogeneity of structures are discussed in
subsection~\ref{aniso:inhom}.

\paragraph{Trion resonant pumping.}

The system of kinetic equations describing spin dynamics of
resident electrons and trions at low pump densities under trion
resonant excitation has the following form \cite{zhukov07}
\begin{eqnarray} \label{system}
&&\frac{d S_z}{dt} = \omega_{\mathrm L} S_y  - \frac{S_z}{\tau_s}
+
\frac{S_T}{\tau^T_0 } - \frac{n_e}{4} G(t)\:, \\
&&\frac{d S_y}{dt} = - \omega_{\mathrm L} S_z  - \frac{S_y}{\tau_s} \:, \nonumber \\
&&\frac{d S_T}{dt} =  - \frac{S_T}{\tau^T}  + \frac{n_e}{4}
G(t)\:. \nonumber
\end{eqnarray}
Here $S_T= (T_+ - T_-)/2$ is the effective spin density of the
trions, $S_y$ and $S_z$ are the electron spin density components.
Function $G(t)$ in Eq.~\eqref{system} describes trion generation
rate. It is assumed that $z$-axis coincides with the growth
direction of the QW structure and the magnetic field is applied
along $x$-axis. The solution for $S_z$, $S_y$ and $S_T$ in
steady-state [$G(t)$ is a constant] readily writes
\begin{eqnarray}\label{sols}
&&S_z = -\frac{n_e}{4} G \frac{\tau_s\tau^T}{\tau_s^T} \frac{1}{1+ (\omega_{\mathrm L} \tau_s)^2},  \\
&&S_y = -S_z \omega_{\mathrm L} \tau_s\, \nonumber\\
&&S_T = \frac{n_e}{4} G \tau^T. \nonumber
\end{eqnarray}
Field dependence of $S_z$ is described by the Lorentz curve, in
agreement with Eq.~\eqref{eq1}. First line of Eq.~\eqref{sols} at
$\omega_{\mathrm L} = 0$ agrees with Eq.~\eqref{low}. In the
absence of trion spin relaxation ($\tau_s^T\to \infty$)
application of an in-plane magnetic field does not result in
steady-state spin polarization. This is in contrast with
time-resolved experiments, where spin beats could be observed even
for $\tau_s^T\to \infty$ (see Fig.~\ref{fig:trion} in
Sec.~\ref{aniso:inhom} and Ref.~\cite{zhukov07}).

Typically, the trion spin relaxation time is much shorter than the
electron spin relaxation time, $\tau_s^T\ll \tau_s$. Therefore, in
moderate magnetic fields $\omega_{\mathrm L} \tau_s\sim 1$ the
steady-state spin of electrons significantly exceeds the trion
spin, $|S_z| \gg |S_T|$.

\paragraph{Exciton resonant pumping.} In this case total spin
of both resident electrons $S_z$ and electrons in excitons $S_{Xz}
= (X_+-X_-)/2$ should be considered. Using Eqs.~\eqref{steady:exc}
in the rotating (with the frequency $\omega_{\rm L}$) frame of
reference, one can derive the following equations describing the
steady state spin of electrons and excitons in the weak pumping
regime ($G_X\tau_s\ll 1$)
\begin{eqnarray}\label{hanle:exc}
\bm \omega_{\rm L} \times \bm S_{X} - \frac{\bm S_X}{T_X} = \frac{1}{2} \bm G_X,\\
\bm \omega_{\bm L} \times \bm S - \frac{\bm S}{\tau_s} + \gamma
\frac{n_e}{2}\bm S_X =0\nonumber.
\end{eqnarray}
Here $\bm S_X$ and $\bm S$ are the vectors of spin of
electrons-in-excitons and of resident electrons, respectively,
$\bm G_X$ is a vector directed along $z$-axis with an absolute
value $G_X$ and $T_X^{-1} = (\tau_0^X)^{-1}+ (\tau_s^X)^{-1} +
\gamma n_e/2$ is the inverse lifetime of electron-in-exciton spin.

The solution of Eq.~\eqref{hanle:exc} for the $z$-component of
spins reads
\begin{eqnarray}\label{hanle:exc:1}
S_{Xz} = -\frac{1}{2}\frac{G_X T_X}{1+ \omega_{\rm L}^2 T_X^2},\\
S_z = \frac{\gamma n_e{ \tau_s} S_{Xz}}{2}\frac{1-\omega_{\rm L}^2
T_X \tau_s}{1+\omega_{\rm L}^2\tau_s^2} \, .\nonumber
\end{eqnarray}
At $\omega_{\rm L}=0$ and $\gamma n_e/2 \gg 1/\tau_0^X,1/\tau_s^X$
last line of Eq.~\eqref{hanle:exc:1} reduces to
Eq.~\eqref{low:exc}. Provided $T_X \ll \tau_s$ electron spin $S_z
= S_z(\omega_{\rm L} =0)/(1+\omega_{\rm L}^2 \tau_s^2)$, i.e. it
is described by the standard Hanle curve [see Eq.~\eqref{eq1}].

\subsubsection{Pulsed excitation}\label{aniso:inhom}

Spin quantum beats described by Eq.~\eqref{read2} are observed in
direct time-resolved experiments under pulsed excitation. Here, we
discuss the initial phase $\varphi$, the precession frequency
$\tilde{\Omega}$ and the spin decay constant ${T}_2^*$ entering
Eq.~\eqref{read2} in detail.

\paragraph{Initial phase.}

\begin{figure}[tbp]
\includegraphics[width=0.35\linewidth]{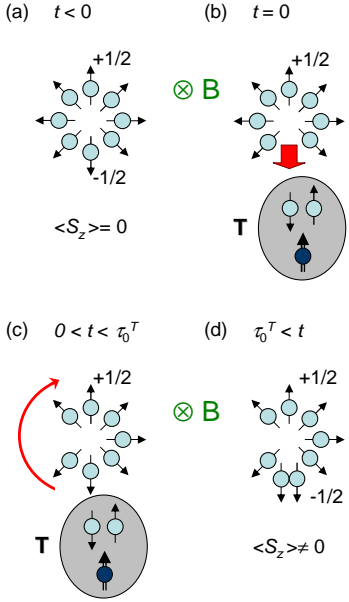}
\caption {Scheme of generation of resident electron spin coherence
in external magnetic fields via resonant photogeneration of
trions. (a) Initial spin state of resident electrons which spin
polarization in the plane perpendicular to the magnetic field is
zero. Spins of resident electrons are precessing around $B$. (b)
$\sigma^-$ polarized photon generates a $(e, +1/2; hh, -3/2)$
electron-hole pair, which captures a $-1/2$ resident electron to
form a trion. The resident electrons become polarized due to
uncompensated $+1/2$ electron spin left. (c) During trion
lifetime, $\tau^T_0$, the net spin polarization precesses around
the magnetic field. Trion state does not precess in magnetic field
as on the one hand its electronic configuration is singlet and on
the other hand in-plane hole g-factor is zero. (d) After trion
recombination the $-1/2$ electron is returned (we neglect here
spin relaxation of the hole in the trion). Final spin state of
resident electrons with induced polarization is
shown.}\label{fig:trion}
\end{figure}

The initial phase of spin beats in the case of trion resonant
excitation strongly depends on magnetic fields. The reason is
schematically illustrated in Fig.~\ref{fig:trion}. After pulsed
excitation some of resident electrons with a certain spin
projection are picked out to form trions. In the absence of hole
spin relaxation the returning electrons after trion radiative
recombination exactly compensate the generated spin of resident
carriers. The transverse magnetic field breaks such a balance: the
spins of electrons in trions do not precess while the spins of
resident electrons rotate around the magnetic field, therefore the
compensation is not complete. Detailed analysis shows that the
electron spin as a function of time can be recast
as~\cite{zhukov07}
\begin{equation} \label{sza}
S_z(t) = \frac{n_0^T}{2} [ |1 - \eta| \sin{(\omega_{\mathrm L} t -
\varphi)} {\rm e}^{ - t/ \tau_s }  - \eta' {\rm e}^{ - t/\tau^T }
]\:,
\end{equation}
where
\begin{equation}\label{phaseNEW}
\varphi = \arctan[{(1- \eta')/\eta''}],
\end{equation}
$\eta'$ and $\eta''$ are the real and imaginary parts of $\eta=
(\tau^T_0)^{-1}/[(\tau^T)^{-1} - \tau_s^{-1} - {\rm i}
\omega_{\mathrm L}]$. Hence, the initial phase is a non-monotonic
function of the magnetic field.

\paragraph{Inhomogeneous distribution of electron $g$-factor.}

In order to take into account inhomogeneous broadening we assume
that electron states are localized and characterized by a
distribution of $g$-factors, $f(g)$. Hopping between different
state is neglected, therefore it suffices to average the single
electron spin precession $\exp{(-t/\tau_s)}\cos{(\omega_{\rm L} t
+ \varphi)}$ over this distribution. In the case of cw-pumping the
Hanle curve, \eqref{sols}, should be averaged over $f(g)$ as well.

It is possible to obtain the analytical results when $f(g)$ is
described by the Lorentzian function
\begin{equation}\label{fdist}
f(g_e) = \frac{\sigma}{\pi [(g_e-g_0)^2 + \sigma^2]},
\end{equation}
where $g_0$ is the average value of $g$-factor and $\sigma$ is the
distribution width. Instead of \eqref{sols} we have for $\langle
S_z \rangle$
\begin{equation}\label{averHanle}
\langle S_z \rangle \sim \frac{1+ \sigma \mu_B B
\tau_s/\hbar}{(1+\sigma \mu_B B \tau_s / \hbar)^2+(g_0 \mu_B B \tau_s /
\hbar)^2}.
\end{equation}

We note that Eq.~\eqref{averHanle} deviates from the standard
Hanle curve in the magnetic fields $B$ where $\sigma \mu_B B
\tau_s / \hbar$ becomes of the order of unity. Typically, in QWs
the $g$-factor spread is small, so that $\sigma \ll g_0$. As a
result, inhomogeneous broadening does not strongly modify the
Hanle curves in the region of $g_e \mu_B B \tau_s / \hbar \sim 1$.

The influence of inhomogeneous distribution of $g$-factors on spin
beats is different. They can still be observed in high magnetic
fields when $g_e\mu_B B \tau_s / \hbar \gg 1$. We obtain averaging
with Eq.~\eqref{fdist}
\begin{equation}
\label{read3} \langle S_z(t)\rangle \sim \exp{(-\sigma \mu_B B t /
\hbar - t/\tau_s)} \cos{(\omega_{\mathrm L} t + \varphi)}.
\end{equation}
Comparison with Eq.~\eqref{read2} suggests that inhomogeneous
broadening of the electron $g$-factor results in additional spin
decay channel characterized by a time constant $T_2^{inh}(B) =
(\sigma \mu_B B / \hbar)^{-1}$. In strong enough magnetic fields
(provided $T_2^{inh}(B) \ll \tau_s$) this spin decay channel
dominates spin beat dephasing process $T_2^* \approx T_2^{inh}$.

\paragraph{Anisotropy effects.} In semiconductor quantum wells
spin relaxation is known to be anisotropic, see
Ref.~\cite{averkiev06} and references therein. In this case the
inverse spin relaxation times are described by a tensor. In general case of an
asymmetric quantum wells there are three linearly-independent
components $\tau_{xx}$, $\tau_{yy}$ and $\tau_{zz}$, where
$x\parallel [1\bar 1 0]$, $y\parallel [110]$ and $z\parallel
[001]$. The spin relaxation times $T_2$ and the precession
frequency $\tilde{\Omega}$ entering Eq.~\eqref{read2} are written
then \cite{kalevich97,glazov08}
\begin{equation} \label{ll12b}
\frac{1}{ T_2 }=  \frac12 \left( \frac{1}{\tau_{zz}} +
\frac{1}{\tau_{yy} } \right)\:,\: \tilde{\Omega} =
\sqrt{\omega_{\mathrm L}^2 - \frac14 \left( \frac{1}{\tau_{zz} } -
\frac{1}{\tau_{yy} } \right)^2}\:.
\end{equation}
Note, that the magnetic field is directed along $x$-axis. The
Hanle effect is described by a Lorentzian ($\sim
[1+(\omega_{\mathrm L}T_H)^2]^{-1}$, where $T_H = \sqrt{\tau_{zz}
\tau_{yy}}$) \cite{averkiev06}. Thus, characteristic times
$\tau_{yy}$ and $\tau_{zz}$ can be found
\begin{eqnarray}
\tau_{zz} = \frac{T_H^2}{T_2} - \sqrt{\frac{T_H^4}{T_2^2} - T_H^2},\\
\tau_{yy} = \frac{T_H^2}{T_2} + \sqrt{\frac{T_H^4}{T_2^2} -
T_H^2}.\nonumber
\end{eqnarray}
Here it is assumed that $\tau_{zz}<\tau_{yy}$, otherwise signs
before square roots should be reversed. Rotating magnetic field in
the QW plane allows to restore all components of the spin
relaxation tensor.

\subsubsection{Resonant spin amplification}\label{RSA_section}

\begin{figure}[tbp]
\includegraphics[width=0.4\textwidth]{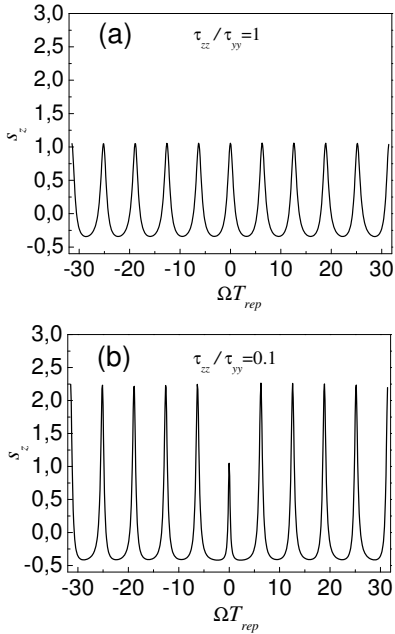}
\caption{
Kerr rotation signals calculated as a function of magnetic field
at a constant pump-probe delay. Panel (a) shows the results in the
case of isotropic spin relaxation ($\tau_{zz}/\tau_{yy}=1$), and panel
(b) corresponds to the strong anisotropy ($\tau_{zz}/\tau_{yy}=1/10$.
Signals are normalized by $s_z(B=0)$. Other parameters are: $T_{\rm
rep}/\tau_{zz}=2/3$, $\Delta t/T_{\rm rep}=1/250$. The details of
calculations are given in Ref.~\cite{glazov08}.}\label{fig:rsat}
\end{figure}

In time resolved experiments the periodic trains of laser pulses
with repetition period of about $T_{\rm rep}=12$~ns are commonly
used. In case when the spin dephasing time is longer that $T_{\rm
rep}$ the signal that does not fully decay from the previous pulse
overlaps with the signal generated by the following pulse.
Depending on the magnetic field strength it may cause constructive
or destructive interference of these contributions, which
complicates evaluation of the dephasing times. To overcome this
complication the technique of the resonant spin amplification has
been suggested~\cite{Kikkawa98}.

In resonant spin amplification (RSA) experiments the magnetic
field dependence of the electron spin is analyzed at a fixed
pump-probe delay. If the pulse repetition period is commensurable
with the spin precession in the magnetic field the spin
polarization is enhanced because the spin injected in the system
at the moments when the precessing spin of resident carriers is
parallel to the spin of photocreated electrons.

The electron spin polarization can be written as
\begin{equation} \label{RSA}
S_z(\Delta t) = \frac{s_0}{2}\ {\rm e}^{ - ( T_{\rm rep} + \Delta
t)/T_2}\ \frac{  {\rm e}^{T_{\rm rep}/T_2} {\cal C}[
\tilde{\Omega} ( T_{\rm rep} + \Delta t) ] - {\cal C}
(\tilde{\Omega} \Delta t)  }{\cosh{(T_{\rm rep}/T_2)} - \cos{
(\tilde{\Omega} T_{\rm rep} )}}\:,
\end{equation}
where $\Delta t\in [-T_{\rm rep},0)$ is the delay between the
probe pulse and the following pump pulse,
\[
{\cal C}(x) = \cos{x} - \frac{1}{2 \tilde{\Omega}} \left(
\frac{1}{\tau_{zz}} - \frac{1}{\tau_{yy}}\right) \sin{x} \:.
\]
According to Eq.~\eqref{RSA} and Fig.~\ref{fig:rsat} the electron
spin polarization represents a series of sharp peaks as a function
of the magnetic field. These peaks correspond to the
commensurability of the spin precession period and the pulse
repetition period. The analysis shows that the ratio of the zero
field maximum to the next maxima (with not too large numbers $N\ll
T_{\rm rep}/|\Delta t|$) is $\eta = (\mathrm e^{T_{\rm rep}/ T_2}
- 1)(\mathrm e^{T_{\rm rep}/\tau_{zz}} - 1)$ \cite{glazov08}. It
equals to $1$ in the case of isotropic spin relaxation, $\tau_{zz}
= \tau_{yy}$, and smaller than $1$ if $\tau_{zz} < \tau_{yy}$.

Inhomogeneous distribution of $g$-factor values can also affect
the resonant spin amplification. As can be shown \cite{glazov08},
the shape of zero-field peak is not affected by  $g$-factor
distribution. However, peaks with larger numbers become wider and
smaller as the effective spin decay constant $T_2^* \approx
T_2^{inh}(B)$ decreases with $B$ [see Eq.~\eqref{read3}].


\section{Experiment}

\subsection{Samples}

We present characteristic experimental results based on two
representative samples of CdTe/(Cd,Mg)Te QWs grown by molecular
beam epitaxy  on  (001)-oriented GaAs substrates
\cite{CdTe_growth}.

\begin{figure}[tbp]
\includegraphics[width=0.6\linewidth]{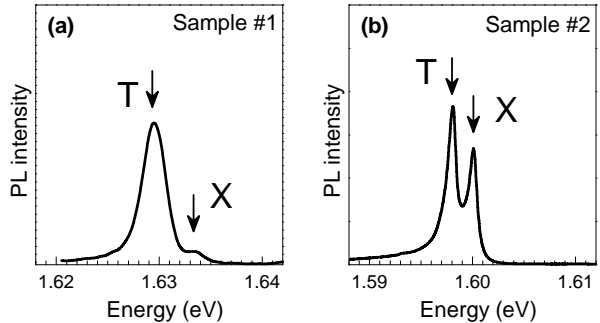}
\caption{(a) PL spectrum of sample \#1. (b) PL spectrum of sample
\#2. The arrows indicate the exciton (X) and trion (T)
transitions. ${T=2}$~K.}\label{fig_PL}
\end{figure}

\paragraph{Sample \#1} is a single 80-{\AA}
$\mathrm{CdTe/Cd_{0.7}Mg_{0.3}Te}$ QW [the PL spectrum is shown in
Fig.~\ref{fig_PL}(a)]. Free electrons in the QW are provided due
to modulation doping by iodine donors in the
$\mathrm{Cd_{0.7}Mg_{0.3}Te}$ barrier at a distance of 100~{\AA}
from the CdTe QW. An electron density in the QW of $n_e = 8 \times
10^{10}$~$\mathrm{cm^{-2}}$ was evaluated by optical means
\cite{3}.

\paragraph{Sample \#2} is a multiple quantum well (MQW) structure
[the PL spectrum is shown in Fig.~\ref{fig_PL}(b)]. It contains
five 200-{\AA}-wide CdTe QWs separated by 1100-{\AA}-thick
$\mathrm{Cd_{0.78}Mg_{0.22}Te}$ barriers. The sample is nominally
undoped. The low concentrations of electrons $n_e = 1.3 \times
10^{10}$~$\mathrm{cm^{-2}}$ in the CdTe QWs are due to residual
\textit{n}-type doping of barriers.

\subsection{Experimental techniques}

\subsubsection{Continuous wave excitation}

For cw pumping, a tunable dye laser is used. Its radiation
wavelength can be set in resonance with either the trion (T) or
exciton (X) states (or one of some other high energy states) of a
QW sample. The excitation is modulated between $\sigma^{+}$ and
$\sigma^{-}$ circular polarizations at a frequency of 50~kHz using
a photoelastic quartz modulator. This allows to avoid nuclear spin
contribution whose spin relaxation is much longer than the
modulation period \cite{Book_OO}.

\paragraph{Hanle effect.} The degree of circular polarization of the photoluminescence
(PL) $P = (I_+^+ - I_+^-)/(I_+^+ + I_+^-)$ is detected by a
Si-based avalanche photodiode and a two-channel photon counter.
PL spectra are dispersed by a 1-m monochromator. The degree of
circular polarization can be selectively detected on either the
trion or exciton emission lines, denoted by $P_T$ and $P_X$,
respectively. When external magnetic field is applied in the QW
plane $P$ is suppressed to zero (Hanle effect). We note, the
circular polarization in emission is sensitive to the
polarizations of resident electrons ($\rho_e$) and excitons
($\rho_X$).

\paragraph{Two-color Hanle-MOKE.} Alternatively, the net spin
polarization of resident electrons can be probed by the
magneto-optical Kerr effect (MOKE) \cite{Kerr1,Kerr2,Hoffmann06}.
In this experiment, one tunable dye laser is used for spin pumping
and another tunable Ti:sapphire laser is used to probe spins. The
latter is linearly polarized, and the photoinduced Kerr rotation
$\theta$ is measured by a balanced diode detector and demodulated
by a lock-in amplifier. Pump and probe energies ($E_{pump}$ and
$E_{probe}$) can be tuned independently, referred to as the two
color mode \cite{Hoffmann06}. By analogy with the Hanle effect,
when external magnetic fields are applied in the QW plane $\theta$
is suppressed to zero. Also note, in the cw pumping regime
concentrations of excitons and trions are much less than the
concentration of resident electrons. Therefore, the photoinduced
Kerr rotation is proportional to the total spin of resident
electrons only, $\theta \propto S_z$ (i.e., contributions from
electrons in excitons and trions can be neglected).

\subsubsection{Pulsed excitation}

Complementary, coherent spin dynamics can be studied by means of a
time-resolved pump-probe Kerr rotation technique \cite{Book_Spin,
Aws02}. It allows direct monitoring the evolution of the spin
coherence of carriers generated by the pump pulses.

\paragraph{Time-resolved Kerr rotation (TRKR).} A Ti:sapphire laser
generates 1.5~ps pulses at a repetition frequency of 75.6~MHz. The
spectral width of pulses is 1.5~meV allowing selectively excite
different resonances in QWs. The laser beam is split in pump and
probe beams (one color mode) and the time delay $t$ between them
is varied by a mechanical delay line. The pump beam is circular
polarized by means of a photoelastic modulator operated at 50~kHz.
The probe beam was linearly polarized, and the Kerr rotation
$\theta$ is measured by a balanced photodetector. From an analysis
of the decay of the Kerr amplitude $\theta (t)$ the spin dephasing
time of electrons $T_2^*$ can be extracted \cite{Zhu06a, Yak07}.
And the precession frequency of $\theta(t)$ allows to determine
the electron $g$-factor.

The TRKR technique can be further improved by using two
synchronized Ti:sapphire lasers \cite{zhukov07}. This allows to
tune independently the pump and probe energies (two-color mode).

Note, for short delay times (in case of pulsed excitation) the
concentration of photocreated excitons can be comparable with the
concentration of resident electrons. Therefore, the Kerr signal is
proportional to the total spin of electrons-in-excitons ($S_X$)
and of resident electrons ($S_z$). This is in contrast to the cw
Kerr signal, where contribution from excitons can be neglected.

\paragraph{Resonant spin amplification (RSA).} It allows
to measure $T_2^*$ times in zero magnetic field limit
\cite{Kikkawa98, Zhu06a}. In this method an external magnetic
field is scanned from small negative ($-20$~mT) to small positive
fields (+20~mT) and a small negative time delay of the probe pulse
is chosen ($\Delta t=-80$~ps). The RSA is very suitable for study
of long living spin beats, when the Kerr signal has considerable
amplitude at negative delays and interferes with the signal at
positive delays.

\subsection{Experimental results: Continuous wave
excitation}

\subsubsection{Hanle effect.} Typical photoluminescence (PL)
spectrum of sample \#1 is presented in Fig.~\ref{fig_PL}(a). It
consists of two lines separated by 3.8~meV. The high-energy line
is attributed to the neutral exciton (X). The low-energy line
originates from the negatively charged trion (T) \cite{4},
consisting of a hole and two electrons in the singlet state
\cite{5}.

\begin{figure}[tbp]
\includegraphics[width=0.6\linewidth]{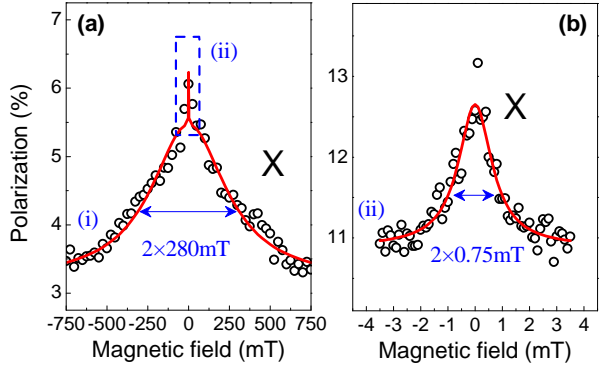}
\caption{(a) A Hanle curve detected at the exciton (X) line of
sample \#1 ($E_{pump} = 1.648$~eV). $W = 2$~$\mathrm{W cm^{-2}}$.
(b) The same rescaled for weaker magnetic fields. $W =
18$~$\mathrm{W cm^{-2}}$. Solid lines are fits to
Eqs.~\eqref{PolXT} and \eqref{hanle:exc:1}. ${T=2}$
~K.}\label{fig1}
\end{figure}

A Hanle curve detected on the exciton line (pump energy $E_{pump}
= 1.648$~eV, $W = 2$~$\mathrm{W cm^{-2}}$) is shown in
Fig.~\ref{fig1}(a). It has two regions: (i) in the high-field
region the polarization decreases slowly from 5.5\% down to 3.5\%,
(ii) in the low-field region $|B| < 3$~mT  a sharp peak is
observed, where the polarization drops from ca. 6\% down to 5.5\%.
The origin of the complex line shape of the Hanle curves in QWs
has been qualitatively understood \cite{2}. The point is that the
trion is formed from the exciton and a resident electron
${X+e\rightarrow T}$, and this formation appears to be spin
dependent. Hence, the degree of circular polarization detected on
the exciton ($P_X$) and trion ($P_T$) emission lines can be
written as \cite{2}
\begin{eqnarray}
&& P_X = \rho_X - \tilde{\gamma} \rho_e ,\label{PolXT}\\
&& P_T=P_X + \rho_e .\nonumber
\end{eqnarray}
Here, $\rho_X \propto S_X + S_{X_h}$ and $\rho_e \propto S_z$ are
spin polarizations of excitons and resident electrons,
respectively, and $\tilde{\gamma}$ is a constant describing the
efficiency of trion formation from an exciton. The total resident
electron and electron-in-exciton spins ($S_z$ and $S_{Xz}$) are
introduced in Sec.~\ref{cwExcitation}. The total hole-in-exciton
spin denoted by $S_{X_h}$ results in incomplete depolarization of
the exciton, as it is not affected by in-plane magnetic fields
(in-plane hole $g$-factor is close to zero). Magnetic field
dependencies of $S_X$ and $S_z$ are described by
Eq.~\eqref{hanle:exc:1}. In the limit $T_X \ll \tau_s$
depolarization of each of them follows the standard Hanle curve of
Eq.~\eqref{eq1} but with different characteristic magnetic field
$B_{1/2}$.

The best fit to Eqs.~\eqref{PolXT} and \eqref{hanle:exc:1} is
shown by the solid line in Fig.~\ref{fig1}(a). From this fit we
find for region (i) $B_{1/2}^X = 280$~mT. It is ascribed to the
depolarization of an electron within the exciton. Due to the
electron-hole exchange interaction (enhanced within the exciton)
the electron spin relaxation time $\tau_s^X$ shortens following
the hole spin flips characterized by a time constant $\tau_{sh}$.
This provides in case of the strong electron-hole exchange
interaction the width of the exciton Hanle curve $B_{1/2}^X =
\sqrt{\tau_{sh} /\tau_0^X} \Delta_0 / |g_e| \mu_B$, where
$\Delta_0$ is the electron-hole exchange splitting \cite{6}.

Regime (ii) in Fig.~\ref{fig1}(a) is ascribed to the
depolarization of resident electrons \cite{2,7}. It is measured
with higher resolution for excitation density $W = 18$~$\mathrm{W
cm^{-2}}$ as presented in Fig.~\ref{fig1}(b). Fitting to
Eq.~\eqref{eq1} (solid line) gives a characteristic magnetic field
$B_{1/2}=0.75$~mT. We find that $B_{1/2}$ depends on the
excitation density in accord with Eq.~\eqref{eq1:W} \cite{7}.
Extrapolation of $B_{1/2}$ to zero excitation density
($W\rightarrow 0$) allows to find the electron spin relaxation
time $\tau_{s}=19$~ns using the electron $g$-factor $|g_e|=1.35$
(deduced from the spin beats as discussed is
Sec.~\ref{ExpPulse_TRKR_1c}).

\subsubsection{Two-color MOKE.}\label{Exp_cw:Two-c}

\begin{figure}[tbp]
\includegraphics[width=0.45\linewidth]{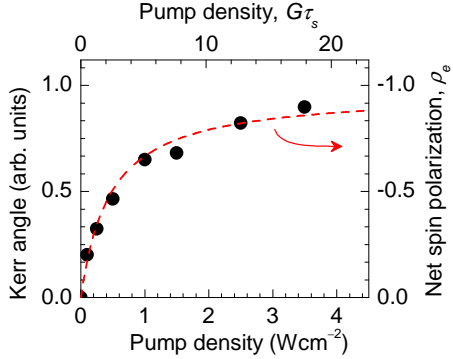}
\caption{Normalized Kerr angle (proportional to the concentration
of pumped electron spins) vs. excitation density for sample \#2
(symbols). Pump ($E_{pump} = 1.598$~eV) and probe ($E_{probe}=
1.597$~eV) energies correspond to the trion resonance. Theoretical
calculation of Eq.~\eqref{gen_spin} is shown by the dashed curve.
Fitting parameters are the same as in Fig.~\ref{fig6}.
${T=2}$~K.}\label{fig2}
\end{figure}

We note that the contribution of the net spin polarization of
resident electrons in the exciton or trion emission may be rather
weak and difficult to detect. As an alternative, the two-color
MOKE offers a highly sensitive technique. We found that the Kerr
angle obtained in cw regime is maximum when the pump energy
coincides or slightly above and the probe energy is  slightly
below the X or T resonance \cite{10}. Hence, independent tuning of
pump and probe energies is an essential requirement for the
two-color MOKE method. A typical dependence of the Kerr angle
$\theta$ on the pump density ${W}$, obtained by this method in
sample \#2 is shown in Fig.~\ref{fig2} \cite{10}. The pump
($E_{pump} = 1.598$~eV) and the probe ($E_{probe} = 1.597$~eV)
energies relate to the trion resonance [PL spectrum of this sample
is presented in Fig.~\ref{fig_PL}(b)]. The concentration of pumped
electron spins $S_z \propto \theta$ increases monotonically with
${W}$ with a tendency to saturation for the pump densities
exceeding 1~$\mathrm{W cm^{-2}}$. Theoretical calculation of
Eq.~\eqref{gen_spin} for the trion resonant pumping [presented in
Fig.~\ref{fig6}(a)] are shown by the dotted curve in
Fig.~\ref{fig2}. It follows quite well the experimental data.

\begin{figure}[tb]
\includegraphics[width=0.5\linewidth]{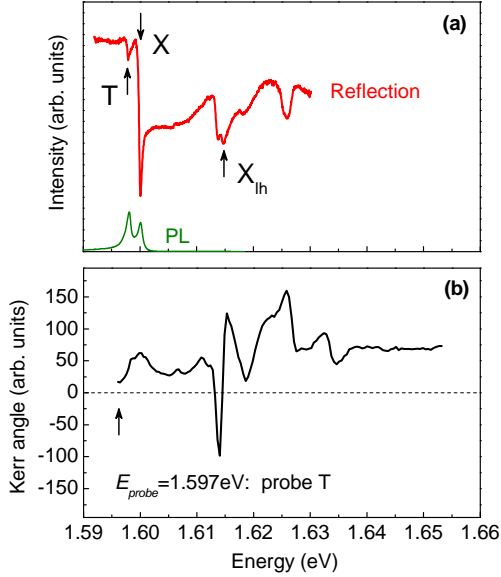}
\caption{(a) PL and reflectivity spectra of sample \#2. The trion
(T), heavy-hole exciton (X) and light-hole exciton
($\mathrm{X_{lh}}$) resonances in the QWs are indicated. (b)
Spectrum of the spin pumping detected by means of the Kerr
rotation at an energy of $E_{probe}=1.597$~eV (slightly below the
trion resonance). $W = 0.25$~$\mathrm{W cm^{-2}}$.
$T=2$~K.}\label{fig3}
\end{figure}

Figure~\ref{fig3}(b) demonstrates efficiency of spin pumping in
sample \#2 ($E_{probe} = 1.597$~eV) as a function of pump energy.
Reflectivity (absorption) spectrum is presented in
Fig.~\ref{fig3}(a) for comparison. A pair of resonances associated
with the neutral (heavy-hole) exciton (X) and negatively charged
trion (T) is well resolved \cite{4,5}. Also several additional
high-energy resonances are clearly seen in the reflectivity
spectrum. Clear correlations in optical spectra of Fig.~\ref{fig3}
are seen. At the low excitation density used ($W = 0.25$~$\mathrm{W
cm^{-2}}$) the efficiency of spin pumping is nearly
independent of whether the pumping is resonant with the heavy-hole
exciton ($E_{pump}=1.600$~eV) or the trion ($E_{pump}=1.598$~eV),
in accord with theoretical consideration in Sec.~\ref{exc}
[Eqs.~\eqref{low} and \eqref{low:exc}]. The net spin polarization
of resident electrons changes its sign when pumping is resonant
with the light-hole exciton ($E_{pump}=1.614$~eV) as expected from
the optical selection rules \cite{Book_OO}. Remarkably, the spin
pumping is still efficient without tendency to decrease for pump
energies up to 50~meV above the bottom of the conduction band,
when electrons with large kinetic energy are created
\cite{GaAs_ExcS}.

\begin{figure}[tb]
\includegraphics[width=0.55\linewidth]{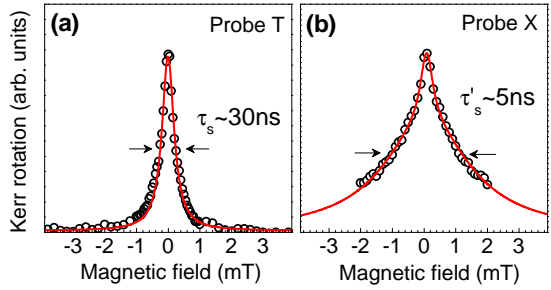}
\caption{The Hanle-MOKE measured on sample \#2 for different probe
energies (a) $E_{probe}=1.597$ ~eV (relates to the trion
resonance) and (b) $E_{probe}=1.599$ ~eV (relates to the exciton
resonance). The pump energy is $E_{pump}=1.600$ ~eV (the exciton
resonance) in both cases. Solid lines are fits (see text for
details). $W = 0.25$~$\mathrm{W cm^{-2}}$. ${T=2}$ ~K.
}\label{fig4}
\end{figure}

Figure~\ref{fig4} shows the Hanle-MOKE curves for different probe
energies while the pump energy is fixed at the exciton resonance,
$E_{pump}=1.600$~eV. These data are recorded at low excitation
(pump density  $W=0.25$~$\mathrm{W cm^{-2}}$). When the probe
energy is set slightly below the trion resonance
$E_{probe}=1.597$~eV the Hanle-MOKE curve is rather narrow
[Fig.~\ref{fig4}(a)]. From the best fit to Eq.~\eqref{sols} [see
also Eq.~\eqref{eq1}] we find $B_{1/2}=0.23$~mT. With the electron
$g$-factor $|g_e|=1.64$ found in spin-flip Raman scattering
experiments \cite{11} or from spin beats
(Sec~\ref{ExpPulse_TRKR_1c}), this characteristic field
corresponds to the spin relaxation time of $\tau_s=30$~ns. This
value is in good agreement with the spin dephasing time
$T_2^*=30$~ns obtained in resonant spin amplification experiments
for the same sample (as discussed in Sec~\ref{ExpPulse_RSA}).

\begin{figure}[tb]
\includegraphics[width=0.39\linewidth]{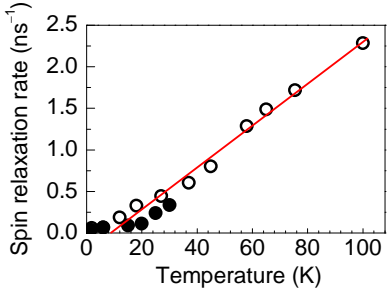}
\caption{Temperature dependence of the spin relaxation rate
obtained in cw experiments ($1 / \tau_s$, solid symbols) and the
spin dephasing rate obtained in pulsed experiments ($1/T_2^*$,
open circles) measured for the pump resonant with the trion energy
in sample \#2. Line is a linear interpolation.}\label{fig:RSA2}
\end{figure}

When the probe energy is tuned to be slightly below the exciton
resonance ($E_{probe}=1.599$~eV), an additional contribution to
the Hanle-MOKE curve appears [Fig.~\ref{fig4}(b)]. We fit these
data (solid line) by a sum of two Lorentz curves of
Eq.~\eqref{eq1} with different $B_{1/2}$ (dashed lines). Details
of fitting procedure are discussed elsewhere \cite{10}. We find
spin relaxation time of $\tau'_s=5$~ns in addition to
$\tau_s=30$~ns. As the spin pump conditions are identical in both
cases [i.e., Figs.~\ref{fig4}(a) and (b)], the difference can be
ascribed only to the selective sensitivity depending on the probe
energy. A possible explanation originates from a spatially
inhomogeneous distribution of resident electrons in the QW plane
\cite{12}. The trions being charged are more sensitive to the
localization (for instance, in the electrostatic potential of
ionized barrier donors) as compared to the neutral excitons. As a
result, when the detection energy is set in resonance with the
trion transition, mostly localized electrons are probed.
Furthermore, the spin relaxation time of localized and free
electrons may be different. Such an explanation is also confirmed
by the temperature dependence of the spin relaxation time detected
when the probe energy is in resonance with the trion state (solid
symbols in Fig.~\ref{fig:RSA2}). With increasing temperature
localized electrons become delocalized, and as a consequence their
spin relaxation rate $1/\tau_s$ rises. In Sec.~\ref{ExpPulse_RSA}
we compare these results with data obtained in TRKR experiments
and discuss  mechanisms of spin relaxation.

\subsection{Experimental results: Pulsed
excitation}\label{ExpPulse}

We turn now to time-resolved experiments performed on the sample
\#2 by means of pump-probe Kerr rotation.

\subsubsection{Time-resolved Kerr rotation (TRKR): One-color mode.}\label{ExpPulse_TRKR_1c}

\begin{figure}[bt]
\includegraphics[width=0.37\linewidth]{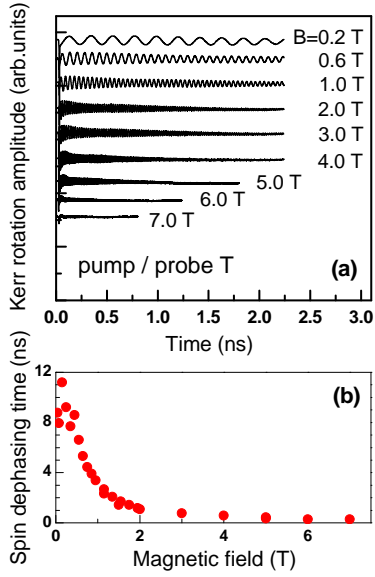}
\caption{(a) Electron spin beats in pump-probe time-resolved Kerr
rotation measures in sample \#2 for different magnetic fields.
Pump and probe are resonant with trion energy. (b) Evaluated spin
dephasing time $T_2^{*}$ vs. magnetic field.
$T=2$~K.}\label{fig:A1}
\end{figure}

Spin beats for different magnetic fields (pump and probe are set
in resonance with the trion) in sample \#2 are shown in
Fig.~\ref{fig:A1}. From fits to Eq.~\eqref{read2} one can find the
spin dephasing time $T_2^*$, the precession frequency
$\tilde{\Omega}$ and the initial phase $\varphi$. In case when
anisotropy effects of Eq.~\eqref{ll12b} can be neglected (high
magnetic fields) $\tilde{\Omega}$ is equal to the Larmor frequency
$\omega_L=g_{{e}}\mu_{{B}}B/\hbar$ allowing to determine precisely
the electron $g$-factor, $|g_e|=1.64$. The spin dephasing time
$T_2^*$ decreases with magnetic field as $1/B$ (see inset) in
accord with Eq.~\eqref{read3}. Hence, in order to avoid the
contribution of inhomogeneous distribution of $g$-factor values to
the spin dephasing process, the spin quantum beats presented below
are measured in the limit of low magnetic fields.

\begin{figure}[tb]
\includegraphics[width=0.37\linewidth]{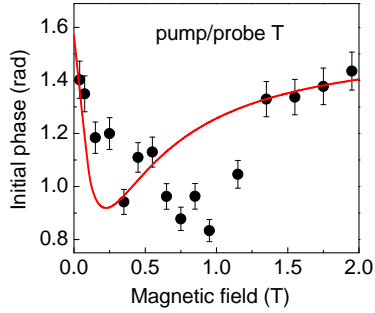}
\caption{Initial phase of the Kerr rotation spin beats versus
magnetic field for the degenerate pump-probe being resonant with
the trion in sample \#2. Pump density is 0.64~$\mathrm{W cm^{-2}}$
and probe density is 0.5~$\mathrm{W cm^{-2}}$. The model
calculation with $\tau_0^T=30$~ps and $\tau_s^T = 60$~ps (solid
line) is performed according to Eq.~\eqref{phaseNEW}. $T=2$~K.
}\label{fig:exp8}
\end{figure}

The initial phase $\varphi$ depends on external magnetic field.
The point is that the spins of the electrons forming a trion are
in a singlet state and therefore do not precess around the
magnetic field as the resident electrons do. Recombination of
trions returns partially $z$-polarized electrons to the 2DEG.
Their spin orientation differs from that of the precessing
electrons, which results in a shift of $\varphi$ \cite{zhukov07}.
We have analyzed the initial phase by extrapolation the spin beats
measured at longer delays back toward zero delay. The results are
given in Fig.~\ref{fig:exp8}. A pronounced minimum is seen at $B
\approx 0.7$~T in qualitative agreement with the model
calculations shown by the solid line [Eq.~\eqref{phaseNEW}],
performed with a trion lifetime $\tau_0^T=30$~ps (obtained from
the PL decay time \cite{zhukov07}) and a spin relaxation time of
the hole in a trion $\tau_s^T = 60$~ps (used as the only fitting
parameter). We also assumed an electron spin dephasing time
$T_2^*=2$~ns (which corresponds to the experimental value at
$B=3$~T). The model demonstrates qualitative agreement with the
experimental data.

\begin{figure}[tb]
\includegraphics[width=0.43\linewidth]{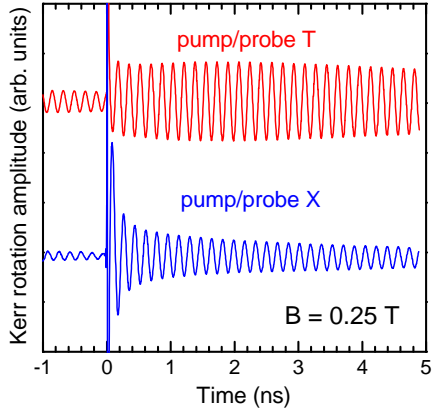}
\caption{Kerr rotation measured by degenerate pump-probe being
resonant either with the trion (upper curve) or the exciton (lower
curve) energies in sample \#2. Pump density is 1.5~$\mathrm{W
cm^{-2}}$ and probe density is 0.3~$\mathrm{W cm^{-2}}$. $T=2$~K.
}\label{fig:exp2}
\end{figure}

Time-resolved Kerr rotation signal  measured for resonant pumping
of either trion or exciton resonances is shown in
Fig.~\ref{fig:exp2}, probe energy coincides with the pump one.
Only long-lived spin beats of resident electrons are observed when
the laser is resonant with the trion and two component decay is
seen for the laser being resonant with the exciton. The
experimental conditions (namely, low pump density 1.5~$\mathrm{W
cm^{-2}}$ and weak magnetic field $B= 0.25$~T) are chosen here to
achieve longer electron spin coherence times \cite{Zhu06a,
zhukov07}. The delay time range between pump and probe in
Fig.~\ref{fig:exp2} covers 6~ns. Under these conditions the spin
dephasing time of the resident electrons reaches 13.7~ns for
pumping in the trion resonance and 4.2~ns for pumping in the
exciton resonance. Moreover the spin coherence does not fully
decay during the time interval of 13.2~ns between the pump pulses
as is clearly seen by the beats at negative delays.

One can see also that the signal measured on the trion resonance
contains only the long-lived spin beats of resident electrons.
However the exciton signal clearly shows two component decay. The
fast decay with a time of 50~ps is due to electron precession in
photogenerated excitons. It disappears with the the radiative
decay of excitons. The long-lived component with a time of 4.2~ns
is due to the resident electrons, whose spin coherence is
generated by excitons captured to trions shortly after
photogeneration.

\begin{figure}[tb]
\includegraphics[width=0.41\linewidth]{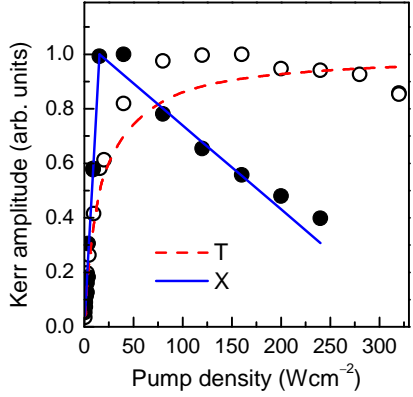}
\caption{Normalized long-lived amplitude of the 2DEG Kerr rotation
measured at a delay of 0.5~ns under resonant pumping of the
excitons (closed circles) and the trions (open circles) in sample
\#2. Model calculations are shown by the lines. The dashed curve
is calculated for trion resonant excitation according to
Eq.~\eqref{trion_res_sat}. The solid lines are calculated for
excitation at the exciton energy according to Eq.~\eqref{se_exc}.
In the latter case, the fitting parameter $\tau_s^X/\tau_0$ is
found to be 10. $T=2$~K.}\label{fig:exp5m}
\end{figure}

Theoretical analysis predicts qualitatively different dependencies
of the Kerr rotation amplitude on excitation density for the
excitons and trions [see Eqs.~\eqref{trion_res_sat},
\eqref{se_exc}]. It is confirmed by experimental data from
Fig.~\ref{fig:exp5m}. Here the normalized values of the Kerr
signals to their maximum values (which, according to our
theoretical predictions, correspond to the electron spin density
being equal to $n_e/4$) are shown. This allows us to compare the
efficiency of spin coherence generation. One can see that in the
low pumping regime the spin coherence generation efficiency per
absorbed photon is practically the same for the laser tuned either
to the exciton or the trion resonance. This is in a good agreement
with cw experiments of Fig.~\ref{fig2} (see also discussion in
Sec.~\ref{Exp_cw:Two-c}) and the theory (Secs.~\ref{tr} and
\ref{xx}): each absorbed photon creates a trion either directly or
via an intermediate excitonic state, thus an electron with a given
spin orientation is removed from the 2DEG. The strong pumping
regime is different for exciton and trion excitation. In the case
of the laser tuned to the trion resonance the spin of the 2DEG
saturates (the small decrease of the Kerr rotation amplitude can
be attributed to heating of the 2DEG) while for the laser tuned to
the exciton resonance a strong decrease of the spin coherence
generation efficiency is seen.

The curves in Fig.~\ref{fig:exp5m} are the results of theoretical
calculations based on the models outlined in Secs.~\ref{tr} and
\ref{xx}. The dashed curve corresponds to trion resonant
excitation, while the solid line is for exciton resonant
excitation. For the dashed curve the only fitting parameter is
the saturation level. For the solid line the only fitting
parameter is the ratio between the electron-in-exciton spin
relaxation time and the exciton radiative lifetime, which is
$\tau_s^X/\tau_0^X=10$. The spin relaxation of an electron in the
exciton $\tau_s^X \sim 0.5$~ns appears to be shorter than that for
the resident electrons. It has been shown that the electron-hole
exchange interaction within the exciton provides an efficient spin
decay channel in external magnetic field \cite{6}.

\subsubsection{Time-resolved Kerr rotation (TRKR): Two-color mode.}\label{ExpPulse_TRKR_2c}

\begin{figure}[tb]
\includegraphics[width=0.43\linewidth]{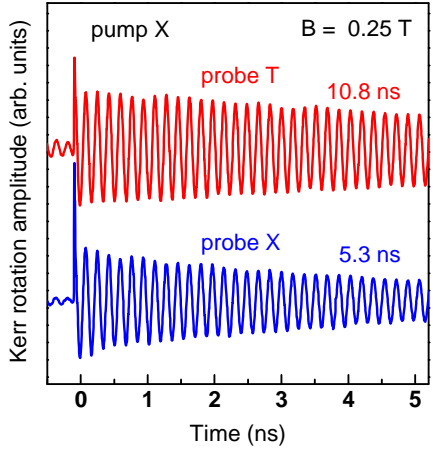}
\caption{Kerr rotation measured by the two-color TRKR in sample
\#2. Probe is resonant with either the trion (upper curve) or the
exciton (lower curve) energies. Pump is resonant with the exciton
in both cases. Pump density is 1~$\mathrm{W cm^{-2}}$ and probe
density is 0.4~$\mathrm{W cm^{-2}}$. $T=2$~K.}\label{fig:exp9}
\end{figure}

In QWs with low electron concentration (as for sample \#2) the
Fermi energy is smaller than the typical localization potential
caused by well width fluctuations. As a result, at low
temperatures a major fraction of electrons is localized. In order
to have a deeper insight into the effects of electron localization
on the spin coherence two-color pump-probe measurements have been
performed in the regime where the longest dephasing times have
been achieved, i.e. using a low pump density in a weak magnetic
field of 0.25~T \cite{zhukov07}. Figure~\ref{fig:exp9} shows the
results of such an experiment, where the pump beam is resonant
with the exciton transition and the Kerr rotation signal is
detected at either the trion or the exciton energy. As the
excitation conditions are identical for both signals in
Fig.~\ref{fig:exp9}, one would expect to detect the same dephasing
times for the 2DEG, irrespective of the probe energy. Therefore,
it is rather surprising to observe that the dephasing times of the
electron coherence differ by a factor of two: $T_2^*=10.8$~ns and
5.3~ns for probing at the trion and the exciton resonance,
respectively.

To explain this difference we suggest that different fractions of
the resident electrons contribute to the Kerr rotation signal
measured at the trion or the exciton energies. This can be
understood if we take into account that different mechanisms lead
to Kerr rotation signal at the exciton and trion energy. For
detection at the trion resonance the effect is contributed by the
variation of the trion oscillator strength, which is directly
proportional to the concentration of electrons with a specific
spin orientation \cite{zhukov07,3}. The trion stability (i.e., its
oscillator strength) increases when resident electrons are
localized in the QW plane \cite{12}, and these electrons possess a
longer spin dephasing time. The Kerr rotation signal detected at
the exciton energy monitors 2DEG spin beats mostly due to the
spin-dependent exciton-electron scattering \cite{Ast00}.
Possibility for this scattering to occur implies existence of free
electrons. Therefore, at the exciton energy we address free or
quasi-free resident electrons which have shorter spin dephasing
times. This result is in good agreement with experimental data and
conclusions drawn from the cw experiments of Fig.~\ref{fig4}.

\subsubsection{Resonant spin amplification.}\label{ExpPulse_RSA}

\begin{figure}[tb]
\includegraphics[width=0.49\linewidth]{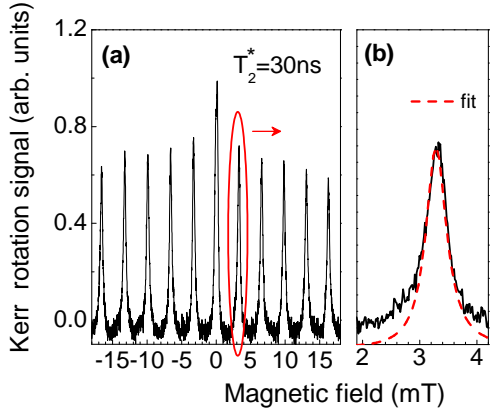}
\caption{(a) Resonant spin amplification measured at a negative
delay of $-80$~ps in sample \#2. Pump density is 0.05~$\mathrm{W
cm^{-2}}$. (b) The fit (dashed line) of a RSA peak to
Eq.~\eqref{RSA}. Spin dephasing time of $T_2^* = 30$~ns is
evaluated. $T=2$~K. }\label{fig:RSA1}
\end{figure}

In certain magnetic fields the spins excited by the pulse train
come in phase resulting in their amplification \cite{Kikkawa98}.
Such resonant spin amplification (RSA) in sample \#2 presented
in Fig.~\ref{fig:RSA1} was obtained for a small negative time delay of
$-80$~ps between the probe and pump pulses. The spin dephasing time
$T_2^*$ time can be directly evaluated from the fit of RSA peak to
Eq.~\eqref{RSA} as it is shown in the inset. At a temperature of
2~K we evaluate $T_2^*=30$~ns, which coincides with the spin
relaxation time $\tau_s = 30$~ns obtained in cw Hanle-MOKE
experiments (Sec.~\ref{Exp_cw:Two-c}). To achieve such a long spin
memory we reduce the pump density to the limit of 0.05~$\mathrm{W
cm^{-2}}$ where the signal has been still detectable.

A temperature increase causes broadening of the peaks, which
reflects shortening of the spin dephasing time \cite{Yak07}. The
similar behavior is observed when the pump density is increased at
a fixed temperature of 2~K. Obviously the effect is related to the
heating of the 2DEG and delocalization of electrons bound to QW
width fluctuations. Free electrons have more channels for spin
relaxation and therefore their spin coherence decays faster.

Temperature dependence of the spin dephasing rate $1/T_2^*$ at
$B=0.25$~T is plotted in Fig.~\ref{fig:RSA2}. Pump density of
0.8~$\mathrm{W cm^{-2}}$ is chosen in order to follow the signal
in the whole temperature range from $T = 2$~K to 100~K. The
dephasing rate is about constant in the temperature range $T = 2 \div
7$~K and it increases linearly with a slope of
0.024~$\mathrm{ns^{-1}K^{-1}}$ up to $T = 100$~K. Characteristic
depth of the electron localizing potential evaluated from the
exciton linewidth in the PL is about 0.5~meV, which corresponds to
an activation temperature of 6~K. For the temperature range from
7~K to 100~K the dependence $1 / T_2^* \propto T$ holds with
high accuracy. Such behavior is characteristic for the
Dyakonov-Perel mechanism of spin relaxation. Theory gives the
following expression for the spin relaxation rate \cite{Dya86,
Book_EL}:
\begin{eqnarray}
\frac{1}{\tau_s}=\beta^2 \tau_p(T)
k_B T \label{eqTaus}.
\end{eqnarray}
Here $\tau_p$ is the momentum relaxation time and $\beta$ is a
coefficient characterizing dependence of the electron spin
splitting on its wave vector. The effects of structure anisotropy
on spin relaxation are neglected. The linear dependence measured
experimentally allows us to suggest that in the studied sample
$\tau_p$ is independent of temperature for $T<100$~K. This is in
agreement with the fact that the electron mobility in CdTe-based
QWs is rather low (usually does not exceed few tens thousands
$\mathrm{V cm^{2}s^{-1}}$) and $\tau_p$ is rather short and falls
in the picosecond range.

\section{Conclusions}

With the use of various techniques based on continuous wave and
pulsed excitation we have performed detailed experimental studies on
optical spin pumping, spin relaxation and spin coherence of
resident electrons in CdTe/(Cd,Mg)Te quantum wells. The
experimental results are substantiated by a theoretical model
based on classical approach to spins. Polarization of the resident
electrons and generation of their spin coherence is provided by a
capture of the resident electrons into trions. The trions in turn
can be either photogenerated by resonant excitation or formed from
the excitons. Variation of pump energy (tuning to resonance with
either the trion or the exciton) highlights details of the
polarization process. Pump density dependencies for two excitation
conditions are nearly the same for low densities (as each absorbed
photon participates in spin pumping) and substantially different
for high densities (for the trion pump the spin polarization tends
to 100\% while for the exciton pump the spin polarization decreases
to zero). Spin beats and spin decay of resident electrons in
external magnetic fields are shown to be subject of spin
relaxation anisotropy and $g$-factor inhomogeneity. Independent
variation of pump and probe energies proves that electron
localization provides an increase of the spin relaxation time, as
also confirmed by temperature dependent experiments. The spin
relaxation time of localized electrons can achieve 30~ns, as
independently found in continuous wave experiment in the limit of
low pump density and in pulsed experiments by means of resonant
spin amplification. Note, this is the longest spin relaxation time
in QW heterostructures reported to date.

\section*{Acknowledgements}

We acknowledge fruitful discussions with E.~L.~Ivchenko. Samples
for these study have been grown in the Institute of Physics,
Warsaw by G. Karczewski, T. Wojtowicz, and J. Kossut. This work
was supported by the Deutsche Forschungsgemeinschaft (SPP1285
Spintronik via grants nos.~YA65/5-1 and OS98/9-1 and by a grant
no.~436~RUS113/958/0-1) as well as  by the Russian Foundation for
Basic Research. MMG was partially supported by the ``Dynasty''
foundation---ICFPM.


\end{document}